\documentclass[apj]{emulateapj}

\usepackage{amsmath}
\usepackage{apjfonts}

\newcommand{\Slash}[1]{\ooalign{\hfil/\hfil\crcr$#1$}}

\def\vector#1{\mbox{\boldmath $#1$}} 

\slugcomment{ApJ in press (\today)}

\shorttitle{Clusters of Galaxies detected by HSC}
\shortauthors{Miyazaki et al.}

\begin{document}
\title{Properties of weak lensing clusters detected on Hyper
  Suprime-Cam's 2.3 deg$^2$ field}
\author{
Satoshi Miyazaki\altaffilmark{1,2}, 
Masamune Oguri\altaffilmark{3,4,5}, 
Takashi Hamana\altaffilmark{1,2},
Masayuki Tanaka\altaffilmark{1}, 
Lance Miller\altaffilmark{6}, \\
Yousuke Utsumi\altaffilmark{7}, 
Yutaka Komiyama\altaffilmark{1,2}, 
Hisanori Furusawa\altaffilmark{1}, 
Junya Sakurai\altaffilmark{2,1}, \\
Satoshi Kawanomoto\altaffilmark{1}, 
Fumiaki Nakata\altaffilmark{8}, 
Fumihiro Uraguchi\altaffilmark{8}, \\
Michitaro Koike\altaffilmark{1}, 
Daigo Tomono\altaffilmark{8}, \\
Robert Lupton\altaffilmark{9}, 
James E. Gunn\altaffilmark{9},
Hiroshi Karoji\altaffilmark{10},\\
Hiroaki Aihara\altaffilmark{5},  
Hitoshi Murayama\altaffilmark{5},
Masahiro Takada\altaffilmark{5}
}
\email{satoshi@subaru.naoj.org}

\altaffiltext{1}{National Astronomical Observatory of Japan, 2-21-1
  Osawa, Mitaka, Tokyo 181-8588, Japan}
\altaffiltext{2}{SOKENDAI (The Graduate University for Advanced
  Studies), Mitaka, Tokyo 181-8588, Japan}
\altaffiltext{3}{Department of Physics, Faculty of Science,
  University of Tokyo, Bunkyo, Tokyo 113-0033,  Japan}
\altaffiltext{4}{Research Center for the Early Universe, University of Tokyo, 7-3-1
 Hongo, Bunkyo-ku, Tokyo 113-0033, Japan}
\altaffiltext{5}{Kavli Institute for the Physics and Mathematics of
 the Universe (Kavli IPMU, WPI), University of Tokyo, 5-1-5
 Kashiwanoha, Kashiwa, Chiba 277-8583, Japan}
\altaffiltext{6}{Department of Physics, Oxford University, Keble Road,Oxford OX1 3RH, United Kingdom}
\altaffiltext{7}{Hiroshima Astrophysical Science Center, Hiroshima
  University, Higashi-Hiroshima, Hiroshima 739-8526, Japan}
\altaffiltext{8}{Subaru Telescope, National Astronomical Observatory
  of Japan, 650 N Aohoku Place Hilo HI96720}
\altaffiltext{9}{Department of Astrophysical Sciences, Princeton
  University, 4 Ivy Ln, Princeton, NJ 08544}
\altaffiltext{10}{National Institutes of Natural Sciences, 4-3-13 
  Toranomon, Minato-ku, Tokyo 105-0001, Japan}

\begin{abstract}
We present 
properties of moderately massive clusters of galaxies detected by 
the newly developed Hyper Suprime-Cam on the Subaru telescope using
weak gravitational lensing. Eight peaks exceeding a S/N ratio 
of 4.5 are identified on the convergence S/N map of a 
2.3 deg$^2$ field observed during the early commissioning phase of the
camera. Multi-color photometric data is used to generate optically
selected clusters using the CAMIRA algorithm. The optical cluster positions were
correlated with the peak positions from the convergence map. 
All eight significant peaks have 
optical counterparts. The velocity dispersion of clusters
are evaluated by adopting the Singular Isothemal Sphere (SIS) fit 
to the tangential shear profiles, yielding virial mass estimates,
$M_{500_{c}}$, of
the clusters which range from 2.7$\times$10$^{13}$ to 
4.4$\times$10$^{14}\;M_{\odot}$. The number of peaks is considerably
larger than the average number expected from $\Lambda$CDM cosmology 
but this is not extremely unlikely if one takes the large sample
variance in the small field into account. We could, however, safely
argue that the peak count strongly favours the recent Planck result
suggesting high $\sigma_8$ value of 0.83. The ratio of stellar mass to the
dark matter halo mass shows a clear decline as the halo mass increases. 
If the gas mass fraction, $f_g$, in halos is universal, as has been suggested
in the literature,  the observed baryon mass in stars and gas shows a
possible deficit compared with the total baryon density estimated from the 
baryon oscillation peaks in anisotropy of the cosmic microwave background. 

\end{abstract}

\keywords{cosmology: observations---dark matter---large scale---structure of universe}

\section{Introduction}
Clusters of galaxies are the largest gravitationally bound systems in
the universe and have been useful cosmological probes to learn about the
geometry and structure of the universe. X-ray observations have, to date,
provided the most efficient way to collect samples of clusters. 
Cluster catalogs compiled from the ROSAT All-Sky Survey \citep{voges99}
have often been used in combination with follow-up observations by
XMM/Chandra to give rigorous cosmological constraints. 
Based on a sample of 238 clusters with redshifts ranging from 0 to 0.5,
\cite{mantz10}
present constraints on the density parameters ($\Omega_{\rm M}$ and
$\sigma_8$) from cluster abundance and on the dark energy
parameters ($\Omega_{\rm X}$ and $w_0$) from the redshift evolution of that
abundance. The mass fraction of host gas, $f_{\rm gas}$, measured at
large radii in the most massive halos is also used to as a
cosmological probe, as it provides a direct measure of the luminosity
distance under the assumption of the universality of the gas fraction
in massive halos; it is expected to scale as
$f_{\rm gas}(z) \propto d(z)^{3/2}$, and is sensitive enough to give
a level of constraint on the parameters comparable to those from other
methods \citep{allen08}. So far, no significant
departure from $\Lambda$CDM model is reported. The eROSITA
satellite is scheduled to launch in 2016 and is expected to 
collect yet more clusters in a more distant redshift range, and is expected
to obtain more stringent constraints \citep{predehl10} .

In the mean time, the clusters themselves draw  astronomers'
interests as a site of galaxy formation and galaxy-galaxy interactions.
Galaxy formation is a process which is accompanied by the conversion of baryons 
into stars in a dark matter halo; the star formation efficiency is critically
dependent on the ability of the gas to cool. In a less massive dark matter
halo, the gas is easily reheated and swept out from the center 
by feedback processes such as supernova explosions.
In a massive halo, the cooling time becomes longer as the halo mass
becomes larger and the virial gas temperature higher, 
and galaxy formation is expected to be suppressed. 
It is also argued that AGN could provide an efficient star-formation 
suppression
feedback mechanism in the higher mass range.
These considerations indicate that there is an optimal halo 
mass where star
formation is most effective around $10^{12}\;
M_{\odot}$ and where the stellar mass to halo mass ratio should peak. 
Statistical estimate based on the abundance  matching
technique generally agree with these naive expectations.
\citep{behroozi10}. 

In terms of individual halos, we now have accumulating
observational evidences that the star formation efficiency actually
drops as the host dark matter halo mass increases in the group to
cluster scale, $> 10^{13}\;M_{\odot}$, \citep{lin03, gonzalez07, 
andreon10, zehavi12, leauthaud12, gonzalez13}. These observational
efforts are also motivated by the desire to measure the {\it total}
baryon mass (gas and stars) to dark matter ratio and to compare with the
universal baryon to dark matter density ratio measured from the 
anisotropy of the cosmic
microwave background. This is, of course, related to the interesting 
''missing baryon problem''
\citep{fukugita98,fukugita03}, and quantitative measurements are important. 

However, the discrepancy among the observations are quite large so
far. On the one hand, 
using 12 galaxy clusters at redshifts around 0.1 with 
$M = 1\sim5 \times 10^{14}\; M_{\odot}$, 
\cite{gonzalez13} argue that the difference between the 
universal value and cluster baryon fractions is
less than the systematic uncertainties associated with the mass
determinations. On the other hand, \cite{leauthaud12} insists clear
shortfall based on the observation of X-ray groups with  
$1\sim7 \times 10^{13}\;M_{\odot}$, in COSMOS field. 

This confusion may arise partly from the variety of halo mass
measurement techniques. \cite{gonzalez13} estimate the halo mass from the X-ray
temperature using the usual virial scaling relation in which they
assume the X-ray gas is in hydrodynamic equilibrium. 
The dynamical method adopted in
\cite{andreon10} assumes that the system is virialized, which many not
be true in the outer region of the clusters.  \cite{leauthaud12}
estimates the mass primarily from X-ray luminosity although this is
calibrated by weak lensing. The chosen methods work best in different
mass ranges, and their systematics almost certainly exist and 
are almost certainly different from technique to technique.
\cite{wu15} reported that the gas fraction is anti-correlated 
with stellar mass fraction in their simulation. This could shift the 
ratio of the stellar mass over the halo mass which introduces another 
complexity in the studies based on the optically or X-ray selected 
clusters. 

In this paper, we employ new approach to sample clusters
and measure the halo mass directly by weak lensing with the fewest
assumptions about the dynamical state of the cluster.
As a part of the commissioning run on the Subaru telescope of 
the newly developed Hyper
Suprime-Cam \citep{miyazaki12} camera, we observed a 2.3 deg$^2$ field in
i-band. We use these observations to locate massive dark matter halos directly
via weak lensing by means of the derived convergence map. 
A deep multi-color catalog is used 
to generate an optically selected
cluster catalog with estimated redshifts using the novel CAMIRA algorithm 
\citep{oguri14}. 
This optical catalog is then correlated with the list of peaks 
in the convergence map. These peaks, if {\it not} coincident with optically
selected clusters, can be spurious, generated
by noise, can correspond to chance coincidences of less massive halos along
the line of sight,  or can corresponds to real clusters with very 
high mass-to-light ratio. 

We chose the Deep Lens Survey (DLS) Field \citep{wittman02} for this
investigation, for which
a deep multi-color galaxy catalog is publicly available, to generate
the optical cluster catalog. 
Based on the shear selected cluster catalogs combined with the
luminosity based catalog, we determine the cluster number count and the
redshift distribution over a wide contiguous area over which 
observational systematic errors can be expected to be minimized,
since the data are taken almost simultaneously over the whole field with
an instrument with a wide field and relatively uniform Point Spread
Function (PSF).
We then determine the stellar mass fraction of individual
shear selected clusters. The individual halo mass, estimated
directly through lensing, ranges nicely between the samples of
\cite{leauthaud12} and \cite{gonzalez13}.

Not all the observing facilities allow this approach. In order to
locate the individual dark matter halos, a high resolution surface mass
density map is necessary, which requires a sufficient number density of
faint background galaxies whose shapes are measured with the necessary
accuracy. 
With a much wider field of view (roughly ten times the area) than the
original Suprime-Cam \citep{miyazaki02b}, the 1.5 degree Hyper Suprime-Cam
 (HSC) field on the 8.2 m Subaru telescope has a crucial advantage for a weak
lensing survey. We have an approved plan to carry out a three layer legacy
survey using Hyper Suprime-Cam \citep{miyazaki13}. Based on the pilot
observations presented here, we attempt to demonstrate the prospects of
the power of the legacy survey. 

\section{Data Analysis}
\subsection{Data set}
DLS Field 2 was observed on February 4th, 2013 Hawaii time. As is
shown in Table\ref{exposureinfo}, two pointing centers are chosen whose
angular distance is 0.75 degree apart. Because the field of
view of HSC is 1.5 degree, we have substantial overlap between two
pointings. The exposure time is either 300 sec
(9 exposures on south, 8 exposures on north) or 150 sec
(4 exposures on north), and the total exposure time of south and north
pointing is 2700 sec and 3000 sec, respectively. A circular dithering
pattern of radius 2 arcmin is adopted around each pointing center to 
fill the gap of CCDs.
The position angle is increased 
by 72$\sim$90 degree between exposures. 

\vspace{0.3cm}
\begin{table}
\caption{Exposure information}
\label{exposureinfo}
\begin{tabular}{lllcc}
\hline
Field & Filter & J2000 & T${\rm_{exp}}$ [sec] & Med. Seeing [''] \\ \hline
South & HSC-i & (139.50, 30.00) &  2700   & 0.58\\
North & HSC-i & (139.50, 30.75) &  3000   & 0.57\\
\hline
\end{tabular}
\end{table}

\centerline{{\vbox{\epsfxsize=8.5cm\epsfbox{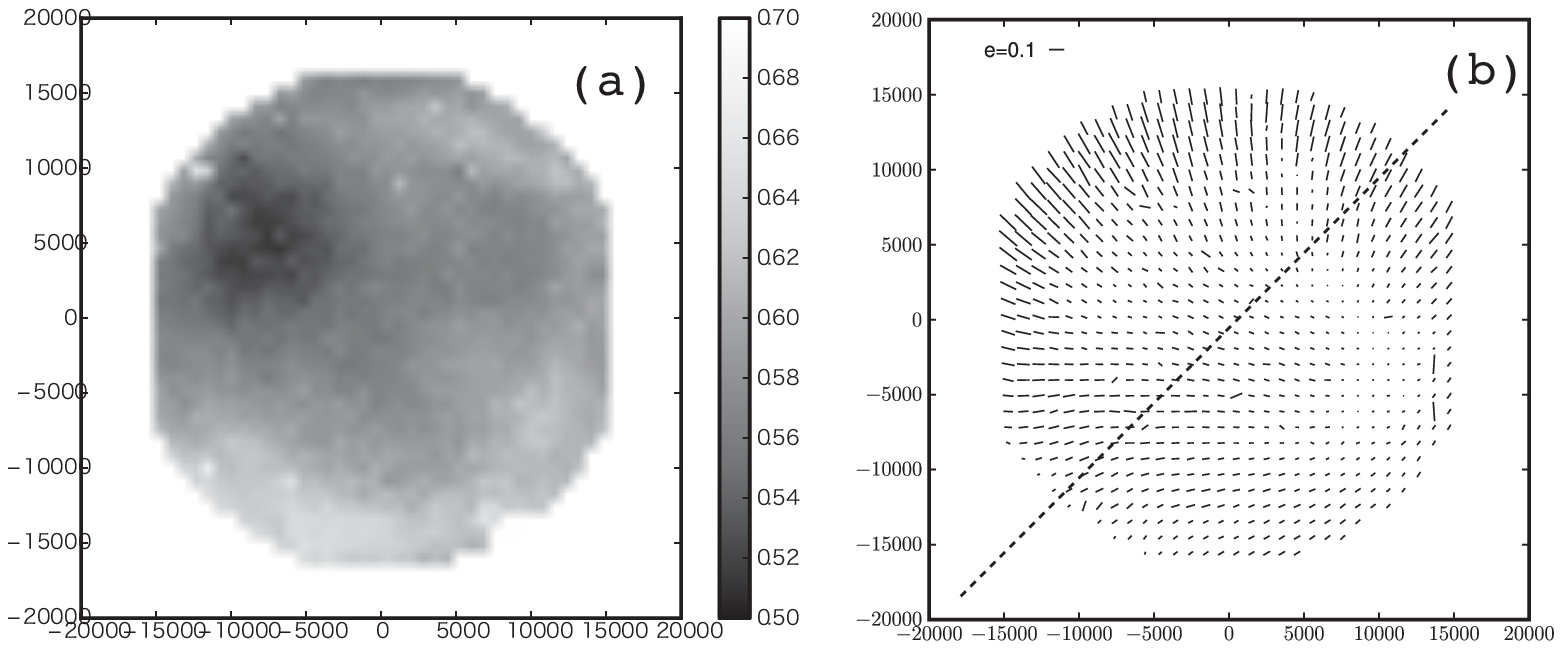}}}}
\figcaption{FWHM map (a) and ellipticity whisker map (b) of one of the
exposure on the south (200 second). On the right, dash line shows the
tilt axis estimated from the ellipticity pattern. The median FWHM and
the ellipticity over the field is 0''.58 and 4.6 \%, respectively.
\label{fig:fwhm_ell}}
\vspace{0.3cm}

The median stellar image (PSF) size of each exposure varied between 0''.52
and 0''.63 and the overall average was 0''.58. Fig.~\ref{fig:fwhm_ell}(a) 
shows the map of the PSF size over the field of view (FOV). The
non-uniformity is visible and we observe a hump of 0''.65 in lower
left bottom and minimum of 0''.51 at slightly upper-right from the
center. This is not expected from the design and the tolerance
analysis and suggests that at this early stage the collimation of the optical
system was insufficiently accurate. 

To investigate this, we evaluate spatial variation of the
ellipticities of stars, $\{e_1, e_2\}$, which is defined as
\begin{equation}
\Slash{e} = \{e1, e2\} = \{I_{11} - I_{22}, 2I_{12}\}/\left(I_{11} +
I_{22} + 2\sqrt{I_{11}I_{22}-I_{12}^2}\right) \label{eqn:ellipticity}
\end{equation}
where $I_{ij}$ is Gaussian-weighted quadrupole moment of the surface
brightness distribution.

Fig.~\ref{fig:fwhm_ell}(b) shows a whisker plot where we present the
position angle of the ellipse with a bar whose length is
proportional to the ellipticity. We find a typical pattern of 
astigmatism generated by the tilt of the camera optics with
respect to the optical axis of the primary mirror. In this case, the
camera is tilted around the axis shown in the figure as a dashed
line. From the start to the end of this observation, the instrument
rotator (InR) rotated by 68 degrees. The fix pattern seen on the FWHM
and whisker plot co-rotated as the InR rotated which supports of our
interpretation of the non-uniformity of the PSF. 

Recall that these test images were taken in the very first engineering
run of the full FOV, we were trying to establish the way to
measure the collimation error. 
Thus these particular images were taken under rather poor alignment
conditions. The median ellipticity over the FOV is roughly 5
\% (Fig.~\ref{fig:fwhm_ell}(b)) which is slightly larger than typical
3$\sim$4 \% raw ellipticity seen on Suprime-Cam images.
(\cite{miyazaki07} Fig.1).
It is hard to determine the inclination angle from these data set
alone but it is roughly a few arcmin level. By the time of this writing,
we have
established a way to measure the tilt and have implementered an auto-focus
system, and we no longer see non-uniform patterns like this \citep{miyazaki14a}.

\subsection{Image Reduction and the galaxy catalog}
HSC has 116 2k$\times$4k CCDs in total: 104 for imaging, in the central
part of the field, and at the edges
4 CCDs for auto guiding and 8 CCDs for focus monitoring. We use the 
auto-guiding 
for the observations reported here. 
Each CCD has 2048$\times$4176 physical pixels 
(15$\mu$m: 0''.169 at the field center). The imaging area of each device
consists of 4 512$\times$4176 stripes;  each stripe is read through an 
independent 
output amplifier located on one side of one of the shorter edges. Each
segment has an 8 pixel wide pre-scan and we add 16
pixels of over-scan along the serial register and 16 lines of
over-scan along the parallel register. 

We use the serial over-scan region for the zero reference of each
amplifier (bias level). The median value of the pixel data on one row is used
for the bias level for pixels 
on that row and subtracted. Non-physical reference pixels are trimmed
from each segment and the segments are tiled together to reconstruct
the original 
2048$\times$4176 image.  We then divide the image by a flat field
image that is generated from an average of dome flatfield images
(typically 10 images). 
Sky level is evaluated on by medianing on a mesh whose element size
is 256$\times$256 pixels and
fitted with 2-D polynomial. This is subtracted from the image. 
This reduced CCD image file is one unit of the data
source for the shape measurement (Section\ref{sec:wlana}).

The basic image analysis pipeline is developed through a collaboration of
multiple institutions including Princeton University, Kavli IPMU and NAOJ.
It is based on a customized version of the ``LSST-stack''
\citep{lsst-stack2,lsst-stack}, a software suite being developed for
the LSST project.
Object detection is made on each CCD image. The PSF on a CCD is modeled as
a function of the CCD pixel coordinates, and can be reconstructed
accurately at any pixel position. 
This is used in the subsequent morphological classification process 
which tags star/galaxy flags as well as flagging cosmic rays. The
flags are stored on the mask layer in an output  FITS file. 

The star catalog is correlated with an external catalog (SDSS-DR8) for
each CCD to perform photometric and astrometric calibration, by using
a cross-matching alogorithm of astrometry.net \citep{astrometry.net}. 
The SDSS magnitudes are transformed into the very similar HSC bands for
photomteric zeropoints, and the astrometry.net is applied to the SDSS
coordinates to determine the world coordinated system of each CCD.


\vspace{0.3cm}
\centerline{{\vbox{\epsfxsize=8.5cm\epsfbox{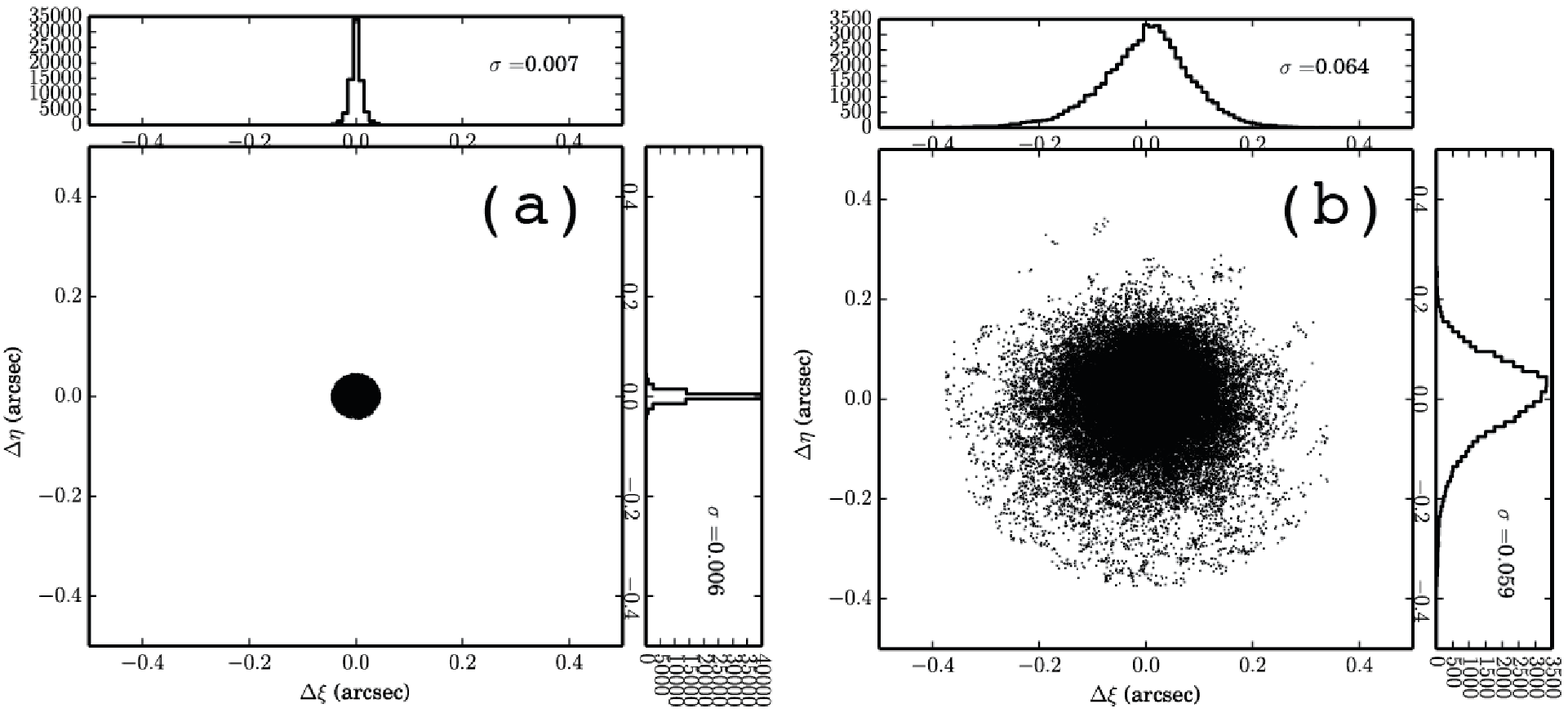}}}}
\figcaption{
The angular displacement of the control stars among exposures in the
mosaic-stack process. The position averaged over the exposures is the
reference (a) whereas the position in the external catalog is the
reference (b). 
\label{fig:mosaicstackerror}}
\vspace{0.3cm}

Because the number of stars on each CCD matched with astrometric
catalog is limited ($\le$ 50), the accuracy of the astrometric solution 
is not sufficient for the later image stacking process. In order to 
minimize the mosaic-stacking error, the residual of i-th control stars 
on a e-th exposure from a reference frame, $(\Delta x_{ie}, \Delta
y_{ie})$, 
is parameterized as a polynomial function
of the field position, $(x, y)$, as 
\begin{equation}
\Delta x_{ie} = \sum^{N}_{l=0}\sum^{l}_{m=0} a_{lme} x^{l-m}y^m \;\;\;\;
\Delta y_{ie} = \sum^{N}_{l=0}\sum^{l}_{m=0} b_{lme} x^{l-m}y^m
\end{equation}
This is a 'Jelly CCD' approach adopted in \cite{miyazaki07}, which is
is particularly important for HSC where we have significant
non-axisymmetric distortion pattern caused by the larger index
mismatch of the glass used for the atmospheric dispersion corrector
\citep{miyazaki14a}. The implementation on the HSC pipeline is more
sophisticated. Rather than dealing with the polynomial as a function
of the pixel coordinate as above, 
the SIP convention (TAN-SIP) of World  Coordinate System (WCS) is
employed to allow the Jelly warp of the CCDs. 
A point on Fig.~\ref{fig:mosaicstackerror} (a)
shows the displacement of the astrometric position of the the control
star with respect to the position on the first exposure, thus the
scatter presents a measure of the mosaic-stacking error. This  internal
error is as low as 6 to 7 mili-arcsec (mas). A point on
Fig.~\ref{fig:mosaicstackerror} (b) shows the displacement from the 
external catalog and the scatter is larger 
(abut 60 mas) which we suppose is partly due to the astrometric error
of the external catalog itself. The peak  position of the distribution
in (b) has no offset from the origin which indicates that there is no
systematic error on astrometric solution. The composition of the internal (a)
and external (b) error gives a estimate of the total astrometric
error of $\sim$ 20 mas in this analysis.
Once the global solution over the multi exposures are obtained, the
WCS of each CCD image is refreshed for further usage in the lensing
analysis.

Handling a entire mosaic-stacked image all at once is usually not
practical. We divide the image into a set of patches
whose size is $4200\times 4200$ pixel. In order to take care of 
objects fallen on the boundary of the patches, we implement 100 pixel
overlap between neighboring patches.  The patch is a unit of 
the following mosaic-stack operation and the object detection on the
co-added image. We then combine the object catalogs of all paces into
a single catalog by removing the objects detected multiple times 
within the overlapped
region. From the combined catalog, we extract the coordinate of 
moderately bright stars (22.0 $<$ HSC-i $<$ 24.5) for star catalog and
the coordinate and the magnitude of galaxies.
The star and the galaxy catalog as well as the image files with
refined WCS of each CCDs (before the stack) are handed over to the next
stage where we carry out the weak lensing shear estimate.

\subsection{Galaxy shape measurements and the shear estimates} \label{sec:wlana}
In the Suprime-Cam weak lensing survey \citep{miyazaki02a,miyazaki07}, 
we employed rather traditional method where we measured the galaxy shapes
on the fully reduced mosaic-stacked CCD image. 
The PSF is not precisely round, nor is it of uniform size over
the whole field.  It is evaluated from star images over the whole field.
(Eqn.\ref{eqn:ellipticity}). 
Because the PSF varies over the field of
view (Fig.~\ref{fig:fwhm_ell}(b)), the PSF ellipticities are
represented as a polynomial function of the field position.
Analysis on the mosaic-stacked image requires that the
PSF pattern does not change much during the series of exposures and
that the CCD boundary does not cause a discontinuity of the PSF
variation. Otherwise, the PSF becomes too complicated to represent as
a simple polynomial function of the field position, and  could cause
systematic error in the galaxy shape measurement.  In the case of
the Suprime-Cam survey, we determined that the conditions were mostly met
because one field was observed in a relatively short time span (40
minutes) with a small ($\sim$ 2 arcmin) dithering offset and we had a
small CCD mosaic. However, in the planned HSC survey one field
observation is split into several semesters to search for variables
and the circular FOV requires relatively large (nearly a half of the
FOV) dithering step. It is therefore not expected that this simple
requirement will be 
met. Instead, the galaxy shapes will be evaluated on the pre-stacked
image and then combined after that to reduce the statistical error. When
the survey is underway, we will develop code to determine shapes and fluxes
simultaneously from the constituent images, which is arguably the
most statistically efficient use of the data.

For this early data,  we have adopted Bayesian galaxy shape measurements
implemented in the ``{\it lens}fit'' algorithm
\citep{miller07,kitching08} 
in this work. The PSF employed in the 
{\it lens}fit is not a simple elliptical but more empirical
32$\times$32 pixel postage stamp image. Each pixel value of the postage
stamp is fitted independently on each exposure to a
2D-polynomial function of the sky-coordinate. 
{\it lens}fit also allows varying low order coefficients 
between CCDs to further minimize the residual on each CCD.

Galaxy shape measurement is made on the 40$\times$40 pixel
postage stamp image of each galaxy. From a galaxy coordinate on the
given catalog, the postage stamp image is trimmed from the un-warped
original CCD image by using the WCS information. Usually, the galaxy
is imaged on multiple exposures and multiple postage stamps are
created for one galaxy. {\it lens}fit has two galaxy model components; 
de Vaucouleurs profile bulge and exponential disk. The center of the
two models are assumed to be aligned. The ellipticity, the 1-D size, the
normalization and the bulge-to-disk ratio are the parameters which
model a galaxy. The postage stamp image of the model galaxy is
convolved with the estimated PSF image at the galaxy position and the
convolved image is compared with observed galaxy image to calculate
the residual. By minimizing the residual, the best fit galaxy model
and the ellipticity is estimated. In the fitting process, 
multiple exposures data are simultaneously fitted.
Note that the galaxy position is
a free parameter as well, and is marginalized by integrating under
the cross-correlation function of the galaxy model and the data.
Therefore, the astrometric error in the mosaic-stack process has
limited impact on the error in the shape measurement. 
Note also that the observed image is not re-sampled nor warped onto
the new coordinate in this process, which avoids the introduction of
correlated noise mentioned in \cite{hamana08}.

For PSF modeling, we employ 2nd order polynominals for the exposure
wide fit and 1st order polynominal for the CCD
term. Fig.~\ref{fig:psfcorrection902032} (a) shows the ellipticity
whisker plot of one exposure that is calculated from the PSF model
postage stamp  and averaged in the grid to visually compare with the
observed PSF shown in
Fig.~\ref{fig:fwhm_ell}. Fig.~\ref{fig:psfcorrection902032} (b) shows
the difference of the observed and modeled ellipticity. The residual
is sufficiently small as $\sigma_e < 1$ \%. 

\vspace{0.3cm}
\centerline{{\vbox{\epsfxsize=8.5cm\epsfbox{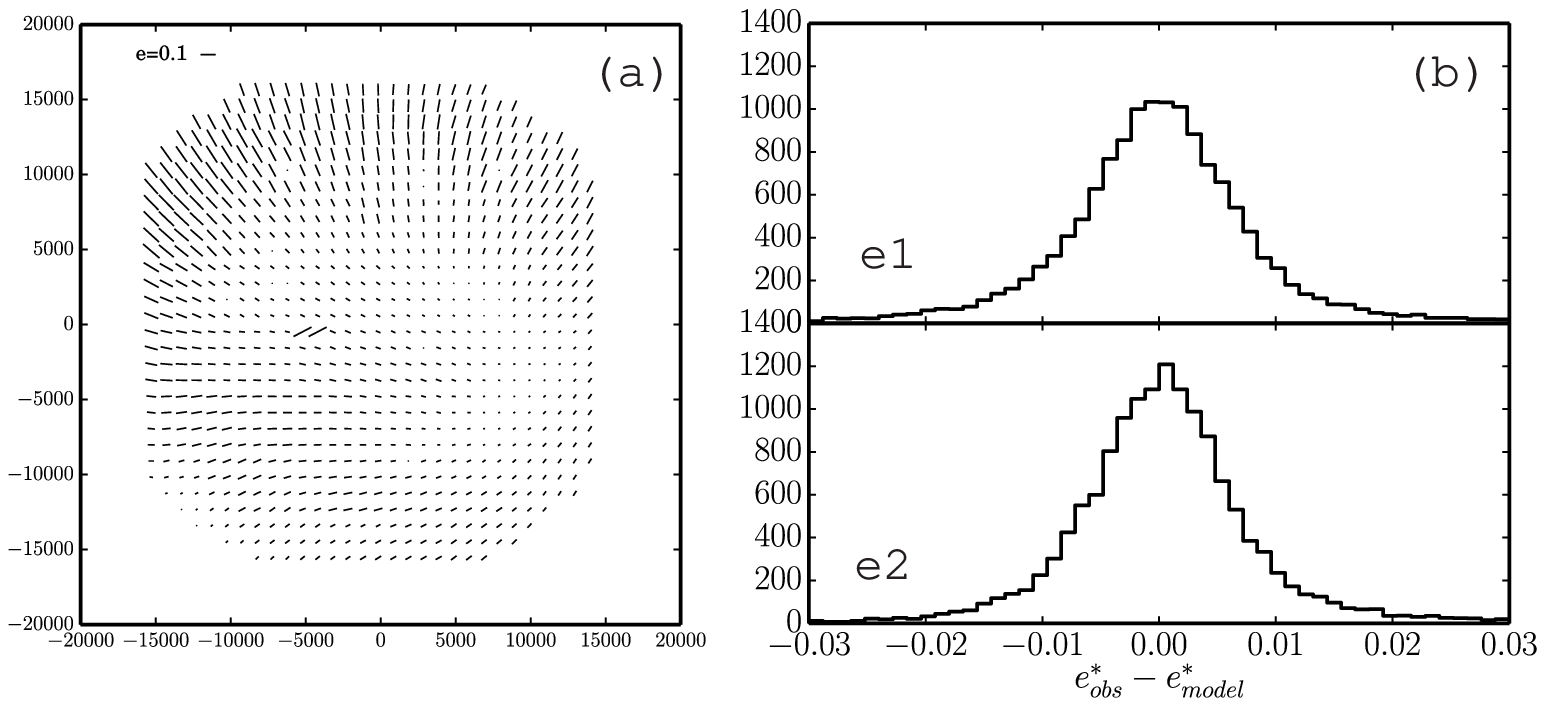}}}}
\figcaption{Ellipticity whisker plot of modeled PSF (a) and the
  residual between the model and the observed data. The sigma of the
  scatter is 0.7 \% and 0.6\%, respectively in each component of the
  ellipticity.
\label{fig:psfcorrection902032}}
\vspace{0.3cm}

In the following sections, we adopt galaxies brigher than $i = 24.5$
for the mass map reconstruction. The rms sigma of the ellipticity
of adopted galaxies is 0.41 and the galaxy number density is 20.9
/arcmin$^2$. In order to estimate the shear bias, we calculate mean
ellipticities of galaxies used for the lensing analysis. The result is
$(\left<e_1\right>, \left<e_2\right>) = (0.005, -0.004)$, which is
sufficiently small for this study. However, the bias is not completely
negligible for cosmic shear studies and further investigations
about the origin will be necessary in the future.

\subsection{Weak Lensing Convergence S/N map}
The dimensionless surface mass density, the convergence
$\kappa(\vector{\theta})$ is evaluated from the tangential shear as
\begin{equation}
\kappa(\vector{\theta}) = \int d^2\phi
\gamma_t(\vector{\phi}:\vector{\theta})Q(|\vector{\phi}|)
\label{eqn:kappa}
\end{equation}
where $\gamma_t(\vector{\phi}:\vector{\theta})$ is the tangential
component of the shear at postion $\vector{\phi}$ relative to the
point $\vector{\theta}$ and  $Q$ is the filter function. We adopt
as a filter the truncated Gaussian 
\begin{equation}
Q_G(\theta) = \frac{1}{\pi\theta^2}\left[ 1 - \left( 1 + \frac{\theta^2}{\theta_G^2}\right)\exp{\left(-\frac{\theta^2}{\theta_G^2}\right)}\right],
\end{equation}
for $\theta < \theta_o$ and $Q_G = 0$ elsewhere \citep{hamana12}. We
employ $\theta_G = 1$ arcmin and $\theta_o = 15$ arcmin, respectively.

We adopt 9$\times$9 arcsec grid cells and then
calculate the convergence, $\kappa(\vector{\theta})$,
on each grid using Eqn.(\ref{eqn:kappa}) 
to obtain the $\kappa$ map. 
In order to estimate the noise of the $\kappa$ map,
we randomized the orientations of the galaxies in the catalog
and created a $\kappa_{noise}$ map. We repeated this randomization
100 times and computed the rms value at each grid point.
Assuming the $\kappa$ error
distribution is Gaussian, this rms represents the 1-$\sigma$ noise
level, and thus the measured signal divided by the rms gives the
signal-to-noise ratio (S/N) of the convergence map at that point. 

\begin{figure*}
\centerline{{\vbox{\epsfxsize=17cm\epsfbox{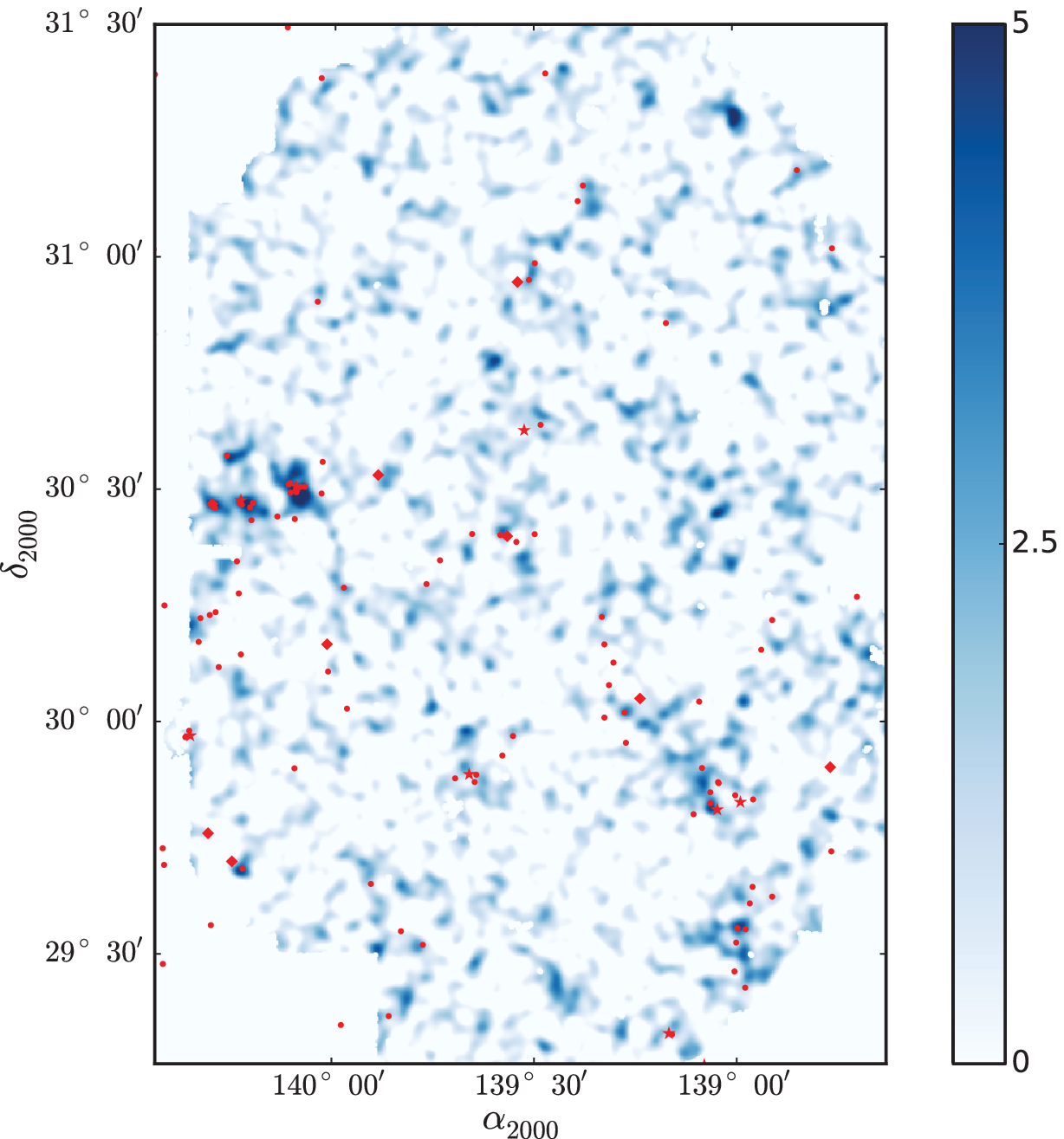}}}}
\figcaption{Weak lensing convergence S/N map reconstructed from the
  shear catalog.
  The smoothing radius $\theta_g$ is 1 arcmin and galaxies
  brighter than 24.5 mag of HSC-i band are employed for the
  reconstruction. Red markers show the  locations of  clusters of
  galaxies registered on NASA/IPAC Extragalactic
  Database where the object type keyword 'GCluster' was used to look
  up. The area below the declination of 31 degree is fully covered by
  Deep Lens Survey where \cite{geller10} carried out the spectroscopic
  campaign. Both Diamond and star markers show the location of SHELS
  clusters which match and un-match with he DLS lensing peaks,
  respectively \citep{geller10}.
\label{fig:kappa.hamana.g2c1.24.5}}
\end{figure*}

Fig.~\ref{fig:kappa.hamana.g2c1.24.5} is the convergence
S/N map so obtained. \cite{wittman06} showed a convergence map based on the DLS
R-band imaging. We find general agreement with that map
although the resolution of the our map is higher. Note, however, that
the convergence map of \cite{wittman06} was generated in the
middle of the survey and it did not employ the full depth image of
DLS. They identifed two shear selected clusters on the map: DLSCL
J0920.1+3029  (140.033, 30.498) and DLSCL J0916.0+2931 (139.000,
29.526) and we see the corresponding peaks on our map. 

The former is a complex Abell 781 cluster region where at least four
clusters have been identified so far from X-ray data and spectroscopic
follow-up observations \citep{sehgal08}. In
Fig.~\ref{fig:kappa.hamana.g2c1.24.5.A781},
we showed a close up view of the region. Crosses indicate the center of
X-ray emissions (XMM) of the four clusters which is called
``West''(z=0.4273), ``Main''(z=0.302), ``Middle''(z=0.291) and
``East''(z=0.4265) respectively from west to east. We detect clear
corespondent peaks in the latter three clusters.  This demonstrates
that the angular resolution of the weak lensing convergence map matches well 
with that of the XMM X-ray image.

We see two separate peaks on the convergence map at the ``Main''
cluster region and the mid-points of two peaks roughly coincides
with the location of the X-ray emission center. 
The X-ray image is mostly round and no corresponding structure 
that is found on the lensing map is observed \citep{sehgal08}. 
\cite{wittman14} presented the convergence map on this A781 
region based on the full depth DLS imaging but no structure 
is seen in the ``Main'' cluster in that map. The structure seen
in Fig.~\ref{fig:kappa.hamana.g2c1.24.5.A781} survived several reality
checks in our analysis (change of the magnitude cut of galaxies and
bootstrap re-sampling of galaxies) but obviously further observations 
are necessary to confirm the structure. If it is confirmed, it will
provide another laboratory for testing the nature of dark matter
following the famous ``bullet cluster''. 
In fact, \cite{sehgal08} reported small extended sub-structure 
to the south-west of X-ray peak. \cite{venturi11} discovered possible
radio relic in their deep radio map. These are indicative of a merger
and already suggested that the ``Main'' is not simple system.

No significant peak is, however, found at the ``West'' position in
Fig.~\ref{fig:kappa.hamana.g2c1.24.5.A781}. In fact, the lensing
signal of the  X-ray emitting cluster Abell 781 ``West'' has been a
matter of debate. Motivated by the less significant detection on the
original DLS map \citep{wittman06}, \cite{cook12} examined the region 
based on images independently obtained  by the Orthogonal Parallel
Transfer Imaging Camera on the WIYN 3.5 m telescope and Suprime-Cam.
They concluded that no significant signal was 
observed on either of the convergence maps and argued that this
example could pose a challenge to the usefulness of
weak lensing for the calibration of
lensing mass against other observable properties of clusters in
observational
cosmology. \cite{wittman14} claimed, however, that 
the strong ``joint constraint'' of \cite{cook12} was an artifact of
incorrectly multiplying p-values. They re-visited the DLS data 
and tried to eliminate the contamination of foreground galaxies
by adopting photometric redshift (photo-z) probability density
weighting in the reconstruction of the convergence map. They suggested
that the peak on the revised convergence map is still less significant
(2.7 sigma level) but this does not strongly exclude the X-ray mass.
We estimate the WL mass at X-ray position of the ``West'' cluster and
the result is shown in Table~\ref{tbl:masscomparison}. It is lower
than what was obtained by \cite{wittman14} but 1-sigma error overlaps 
each other. Therefore, our result supports the summary by
\cite{wittman14} saying that the mass of ``West'' cluster ($M_{200}$)
ranges 1--3 $\times 10^{14}M_{\odot}$.

\vspace{0.3cm}
\centerline{{\vbox{\epsfxsize=8.5cm\epsfbox{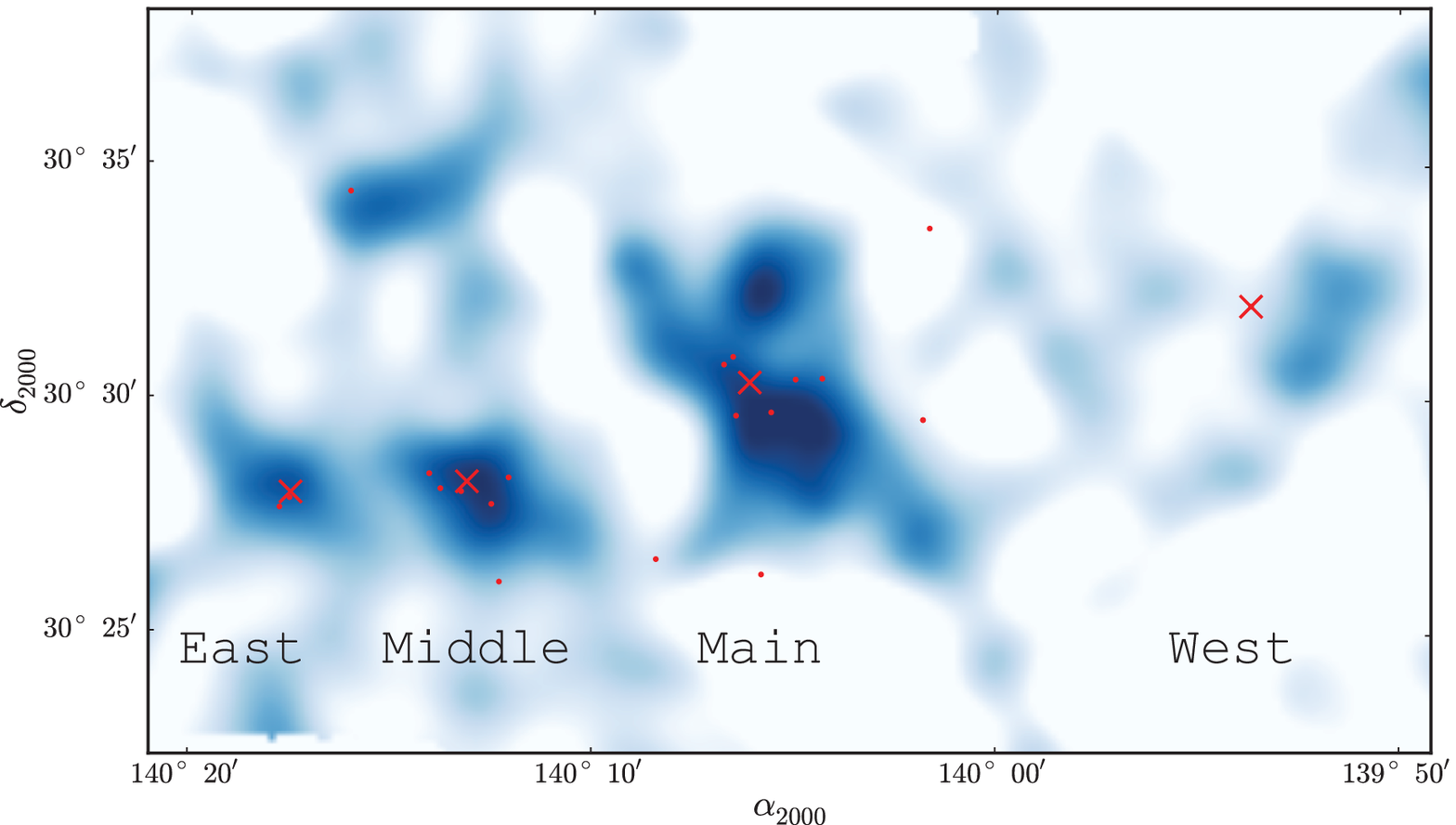}}}}
\figcaption{Close up view of the convergence map
  (Fig.~\ref{fig:kappa.hamana.g2c1.24.5}) at the Abell 781 multi-cluster
  region. Four cross marks are superimposed that indicate the location
  of the X-ray clusters observed by XMM: from west to east ``West'',
  ``Main'', ``Middle''  and ``East''.
\label{fig:kappa.hamana.g2c1.24.5.A781}}
\vspace{0.3cm}

\subsection{Cluster Searches based on the Multi-Color Catalog}
The red symbols on Fig.~\ref{fig:kappa.hamana.g2c1.24.5} show the location
of clusters of galaxies registered on NASA/IPAC Extragalactic database
(NED). It is clear that the symbols tend to exist in the
colored area (stronger lensing signal) and we see the general
corespondence of the weak-lensing peaks and the cluster
positions. Among them, the star and diamond symbols are SHELS
clusters which are identified by \cite{geller10} employing uniform
spectroscopic observation of a magnitude-limited (R < 20.6) sample
using HectoSpec. \cite{geller10} matched their clusters 
with the DLS lensing peaks; the star symbols show the matched
clusters and the diamonds, un-matched. 

In order to make an independent comparison between the light and mass
on this region,
we search for clusters using the DLS public photometric data with
a method, the ``Cluster finding algorithm based on Multiband
Identification of Red-sequence gAlaxies" (CAMIRA), developed by Oguri
(2014). CAMIRA makes use of the stellar population synthesis (SPS) models
of Bruzual \&
Charlot (2003) to compute SEDs of red-sequence galaxies, estimates
the likelihood of being cluster member galaxies for each redshift
using $\chi^2$ of the SED fitting, constructs a three-dimensional
richness map using a compensated spatial filter, and identifies
cluster candidates from peaks of the richness map. For each cluster
candidate, the brightest cluster galaxy candidate is identified based
on stellar mass and location. Readers are referred to Oguri (2014)
for more details of the algorithm and its performance studied in
comparison with X-ray and gravitational lensing data.

We apply the CAMIRA algorithm to the DLS $BVRz$-band data. We apply
a magnitude cut of $R<24.5$ to exclude galaxies with large
photometric errors. In Oguri (2014), a number of spectroscopic
galaxies in SDSS have been used to calibrate the SPS model. We adopt
this SDSS calibration result, but we also add a constant 0.02~mag
error quadratically to the model scatter, in order to accommodate the
systematic offset of magnitude zeropoints and the difference in
magnitude measurements between DLS and SDSS. We confirm that this
SDSS-calibrated SPS model provides reasonable $\chi^2$ values when
fitted to spectroscopic SDSS red-sequence galaxies in the DLS
field. The cluster catalog contains, for each cluster, the cluster center
based on the brightest cluster galaxy identification, the photometric
redshift, and the richness. The redshft and richness ranges are
restricted to $0.1<z<0.8$ and $N>10$, respectively. In addition, in
this paper we also compute the total stellar mass by summing up
weighted stellar mass estimates of individual galaxies; the weights
are the ``weight factors'' $w_{\rm mem}$, which resemble the
membership probability of each galaxy (see Oguri 2014). Stellar mass
estimates of individual galaxies are obtained from the SPS model
fitting in which we assume a Salpeter initial mass function. 

We estimate the stellar mass only using optical data set which could
cause the significant error in the estimation because of the limited
band-pass. When we argue the cluster members in this work, however, we
mainly deal with passvie galaxies whose stellar mass is expected to be
easier to estimate compared with that
of other general galaxies. In fact, \cite{annunziatella14} 
argued that the stellar mass of passive galaxies estimated
from optical data agree well with what is estimated from optical and
NIR data within 25 \% if they adopt the template of passive galaxies in
the calculation. Therefore we expect that our stellar mass estimates are not
significantly biased by the absence of NIR data.

In Fig.~\ref{fig:oguri.new.sn3.7.richnesover10.rl}, we show the
positions of resultant optically selected clusters as open circles where
the center of the clusters is defined as the position of the bright
cluster galaxy (BCG). The redshift is color-coded as is presented on
the side bar. The diameter of the circles is 3 arcmin. 

\section{Results and Discussions}
\subsection{Correlation of the Peaks on the Convergence Map and the
  Optically Selected Clusters}
We have searched for peaks in the convergence map; the results are
indicated on Fig.~\ref{fig:oguri.new.sn3.7.richnesover10.rl} as filled
triangles (significant peaks of S/N > 4.5) and filled squares
(moderate peaks of 3.7 < S/N < 4.5). We then match the peaks with
the centers of the optically selected clusters and count the
coincidences as we vary the 
match tolerance. As is shown in Fig.~\ref{fig:nmatch}, when we increase
the tolerance from zero, the number of matches rises rapidly and
reaches a plateau at about 1.5 arcmin.
The displacement of the BCG from the cluster center (as defined by the 
X-ray emission
center) can be as large as 0.5 Mpc/h \citep{oguri14}, which corresponds
to $\sim$ 1.5 arcmin at the redshift of 0.7; the highest redshift in our
sample. We would therefore expect to have to use a match tolerance of
this order to recover real matches, and this is what we see; at
approximately this tolerance we have recovered all the real matches and
the number should plateau, as it does. 
The number of matches increases again slowly beyond $\sim$ 4
arcmin which we understand as accidental coincidences.

Table\ref{tab:shear_selected_cluster} shows the list of our
peaks sorted by the S/N of the convergence map. 
Among nine significant peaks of S/N > 4.5, five peaks (Peak ID 1, 3,
5, 6, 7) do not have a corresponding optically selected cluster using
the CAMIRA algorithm described in the previous section. We carefully
examine each case here.

To complement our CAMIRA catalog, we looked for the clusters of
galaxies in the NASA/NED database; the matched clusters are shown at
the last
column of the table where the tolerance is set at 1.5 arcmin.
Peak ID 3 and 7 do have counterparts on NED (SHELS
J0920.4+3030:z=0.3004 and WHL J092104.1+303424:z=0.2758,
respectively). These are proxy of the prominent A781 Main cluster (ID
0) and have
similar redshift with the main cluster. 
In the CAMIRA algorithm, we eliminate the
member galaxies of detected clusters to avoid double-counting. CAMIRA
also uses a compensated spatial filter, which suppresses detection
of clusters near very massive clusters. These effects would
explain why the CAMIRA catalog failed to detect clusters at the
position of shear peak ID 3 and 7. The situation might be improved by
modifying the form of the spatial filter for optical cluster finding
but we leave this fine-tuning for the future work. 

Peak ID 5 matched with the original DLS shear selected cluster; 
DLSCL J0916.0+2931 \citep{wittman06} which is another complex system in this
region. They reported that there are three associated X-ray
peaks along north-south line and the north and the south peak were
confirmed spectroscopically as clusters at a redshift of 0.53. 
We, in fact, note that below the richness threshold of 10 adopted
here, there is a optically selected cluster at Photo-z of 0.542
(richness = 8.05) 1.5 arcmin north of the Peak ID 5. The central X-ray
peak is later confirmed as a cluster at z=0.163 \citep{geller10}. 

Peak ID 6 is supposed be a possible sub-structure of the peak ID 0 because
the X-ray emission peaks at just between Peak 0 and Peak 6
(Fig.~\ref{fig:kappa.hamana.g2c1.24.5.A781}). 

We therefore conclude that except peak ID 1 where no deep multi-color data
is available because it is outside of DLS field, all the other
very significant peaks of S/N > 4.5 are generated by physical entities.

At a somewhat lower S/N level, 
\cite{miyazaki07} identified 17 peaks with S/N over 3.7 in a 
2.8 deg$^2$ region in XMM-LSS field,
and found that nearly 80
percent of the peaks have physical counterparts. On the other hand
we have 26 peaks on the DLS overlapped 2 deg$^2$ region and only
50 percent of the moderately significant (S/N > 3.7) peaks  have
identified physical counterparts. The
discrepancy might be partially explained by the different noise level
caused by the conservative magnitude cut adopted in this work which in
turn resulted  in a smaller number density of weak lensing galaxies,
which in turn raises the noise level on the convergence map. When we raise the
threshold from 3.7 to 4.5 we see better match on the DLS field; nine
peaks out of ten have counterparts. Note that number density of the
matched peaks are quite similar on XMM-LSS (4.3 peaks/deg$^2$ for
for S/N 3.7) and DLS (4.0 peaks/deg$^2$ for S/N > 4.5). 

\vspace{0.3cm}
\centerline{{\vbox{\epsfxsize=8.5cm\epsfbox{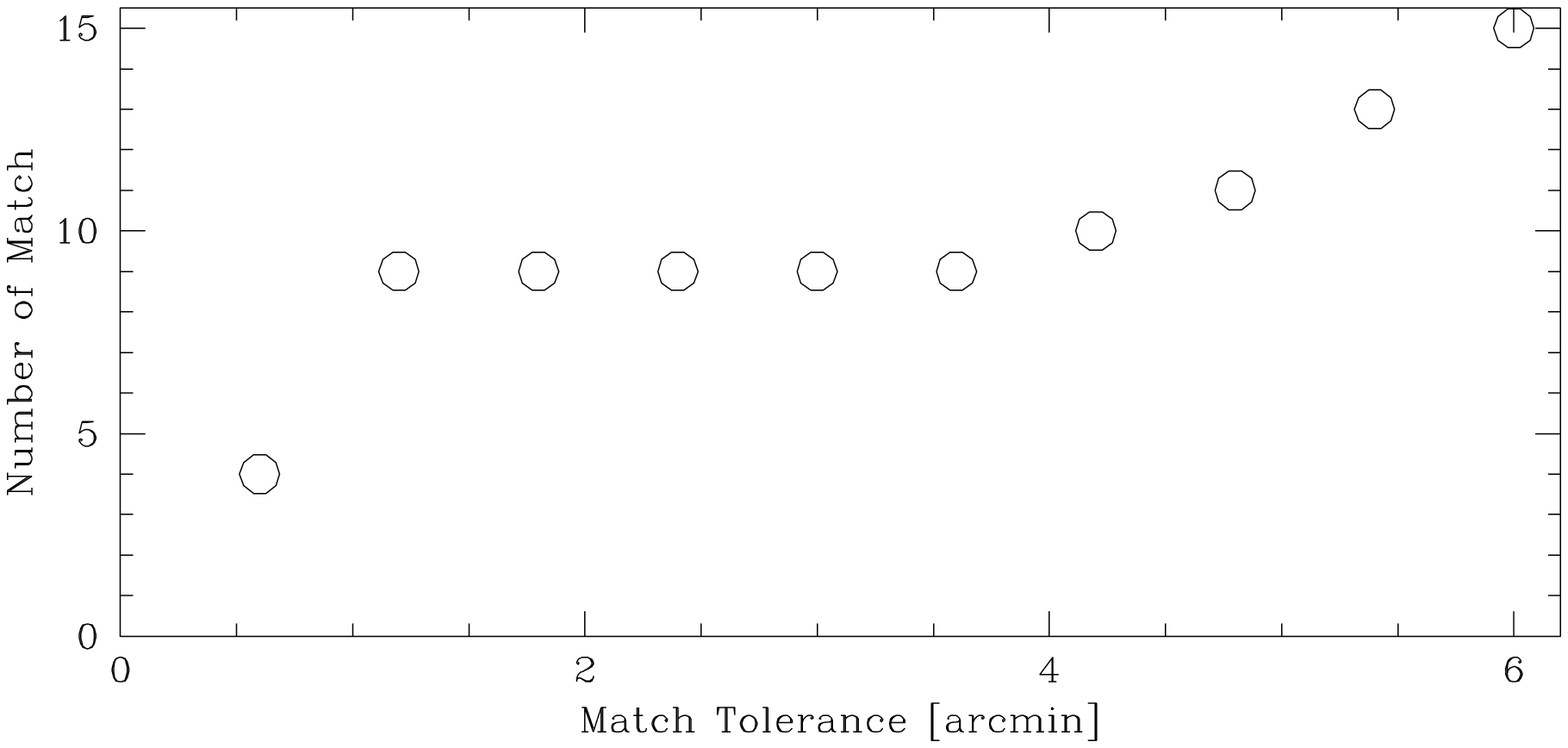}}}}
\figcaption{ 
Number of shear selected clusters matched with optically selected clusters
as a function of the tolerance of the angular distance used for the match.
The the number of matches increases with the tolerance and 
reaches plateau around 1.5 arcmin
and then gradually increases again, perhaps because of accidental matches. 
We set the tolerance at 2 arcmin (9 matches).
\label{fig:nmatch}}
\vspace{0.3cm}

\begin{table*}
\caption{Shear selected cluster samples generated by matching with
the CAMIRA catalog  and external catalogs. 
\label{tab:shear_selected_cluster}}
\begin{center}
\begin{footnotesize}
\begin{tabular}{lccccclll}
\tableline\tableline\noalign{\smallskip}
 ID & S/N    &   RA2000  & DEC2000 &  Photo-z & log(M$_s$) & Richness & Match \\\tableline
 0  & 6.0    &   140.082 & 30.4898 &  0.2807  & 13.046 & 96.872   & SHELS J0920.9+3029 (z=0.2915 A781 Main)\\
 1  & 5.9    &   138.998 & 31.2972 &          &        &          & Out of DLS Field\\
 2  & 5.7    &   139.043 & 30.4525 &  0.6341  & 12.594 & 34.359   & Rank 2\tablenotemark{d}\\
 3  & 5.6    &   140.215 & 30.4691 &  0.3004\tablenotemark{1} & & & SHELS J0920.4+3030 (z=0.3004 A781 Middle)\\
 4  & 5.4    &   139.064 & 29.8176 &  0.514   & 12.738 & 38.957   & SHELS J0916.2+2949 (z=0.5343)\\
 5  & 4.9    &   138.993 & 29.5623 &  0.531\tablenotemark{2}  & & & DLSCL J0916.0+2931 Rank 1\tablenotemark{d}\\
 6  & 4.9    &   140.099 & 30.5372 &  0.2807  &        &          & Possible substructure of Peak ID 0\\
 7  & 4.9    &   140.256 & 30.5689 &  0.2758\tablenotemark{3} & & & WHL J092104.1+303424\\
 8  & 4.8    &   138.982 & 30.0472 &  0.5204  & 12.255 & 12.372   & Rank 3\tablenotemark{4}\\
 10 & 4.4    &   140.299 & 30.4687 &  0.4022  & 12.685 & 41.929   & SHELS J0921.2+3028 (z=0.4265 A781 East)\\
 20 & 3.9    &   139.270 & 30.0231 &  0.3014  & 12.583 & 32.332   & WHL J091705.9+300118 (Photo-z=0.3285) Rank 0\tablenotemark{4} \\
 21 & 3.9    &   139.789 & 29.5207 &  0.326   & 12.276 & 10.285   & WHL J091906.0+293119 (Photo-z=0.3576)\\
 22 & 3.8    &   139.648 & 29.4809 &  0.544   & 12.471 & 11.963   & \\
 27 & 3.7    &   140.224 & 29.6816 &  0.274   & 12.494 & 26.238   & SHELS J0921.0+2942 (z=0.2964)\\\tableline\tableline
\end{tabular}
\tablenotetext{1}{Geller et al.(2010)}
\tablenotetext{2}{\cite{wittman06}}
\tablenotetext{3}{\cite{hao10}}
\tablenotetext{4}{\cite{utsumi14}}
\end{footnotesize}
\end{center}
\end{table*}

\vspace{0.3cm}
\centerline{{\vbox{\epsfxsize=8.5cm\epsfbox{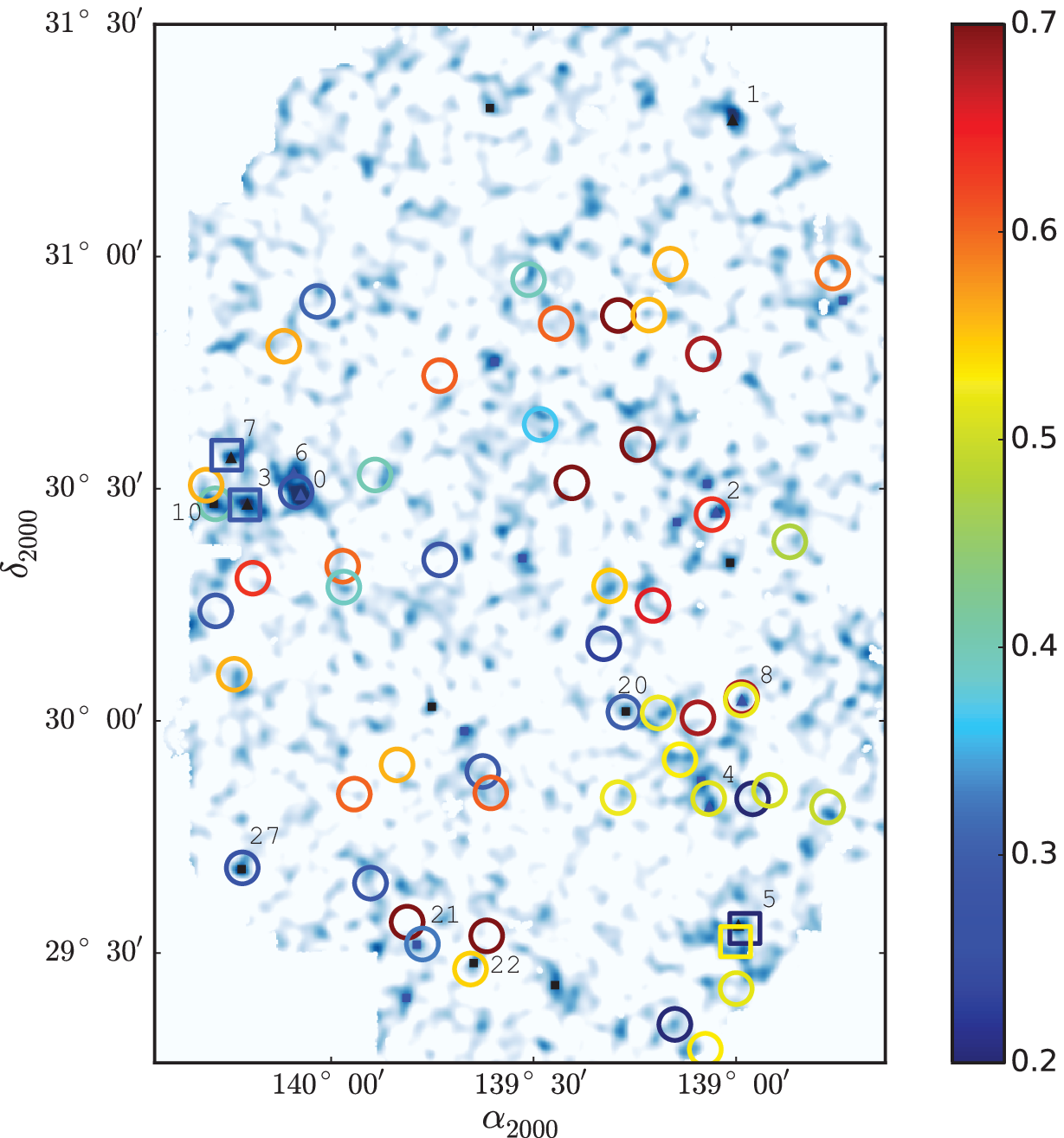}}}}
\figcaption{Peak location  on the convergence map: filled triangle
  (S/N > 4.5) and filled square (3.7 < S/N < 4.5). Open circles show
  the location of of optical clusters identified with the  CAMIRA
  algorithm \citep{oguri14}  whose richness is over 10 and open
  squares are clusters registered on NED (from west to east:
  CXOU J091554+293316 (z=0.184), SHELS J0920.9+3029 (z= 0.291) and
  WHL J092104.1+303424 (z=0.2758). The diameter of the open circles is 3 arcmin.
\label{fig:oguri.new.sn3.7.richnesover10.rl}}
\vspace{0.3cm}

We now examine how the optically selected clusters match with the peaks
on the convergence map in a different way. In Fig.~\ref{fig:z_richness},
circles show the redshift and the richness of the clusters detected by
CAMIRA algorithm. The clusters matched with the most
significant peaks (S/N $\ge$ 4.5) are marked with triangle symbols 
whereas the ones
matched with moderate peaks (3.7 < S/N < 4.5) are marked with squares. It is
encouraging to know that all luminous clusters samples (richness > 32) have
corresponding convergence peaks. 

A cluster (photo-z = 0.29) that is just below the richness threshold
and has no associated peak is  
SHELS J0918.6+2953 whose spectroscopically confirmed redshift is
0.3178. We notice that this is one of the shear selected samples 
in \cite{kubo09} (Rank=5 $\nu$ = 3.9). Although we have positive
convergence signal ($\sim$ 2.5) here we see no strong peak (S/N >
3.7). Also, we have a peak (S/N = 3.9) 5 arcmin north of SHELS
J0918.6+2953. When we adopt a larger smoothing kernel ($\theta_g$ =
2 arcmin) on the convergence map, the positive signal is connected
with the northern peak and results in more significant peak ($\sim$
3.7) which can be a counterpart of SHELS J0918.6+2953. 
In this paper, however, we will keep a single smoothing scale of 1
arcmin for easy comparison with theoretical expectations. 

There is a cluster at photo-z of 0.5204 whose richness is low (12)
but is matched with a significant peak (ID = 8). This peak is reported
in \cite{utsumi14} (Rank 3) as well using a totally independent data set
and data analysis pipeline; Suprime-Cam data analyzed by
imcat. We therefore suppose that this is not a spurious peak caused 
by systematic errors. Although they observe a spatial concentration
of eight galaxies at redshift 0.537 ($\Delta$z = 0.025), they did
not identify the peak as a cluster because it did not meet the
more stringent SHELS cluster criteria \citep{geller10}.

\vspace{0.3cm}
\centerline{{\vbox{\epsfxsize=8.5cm\epsfbox{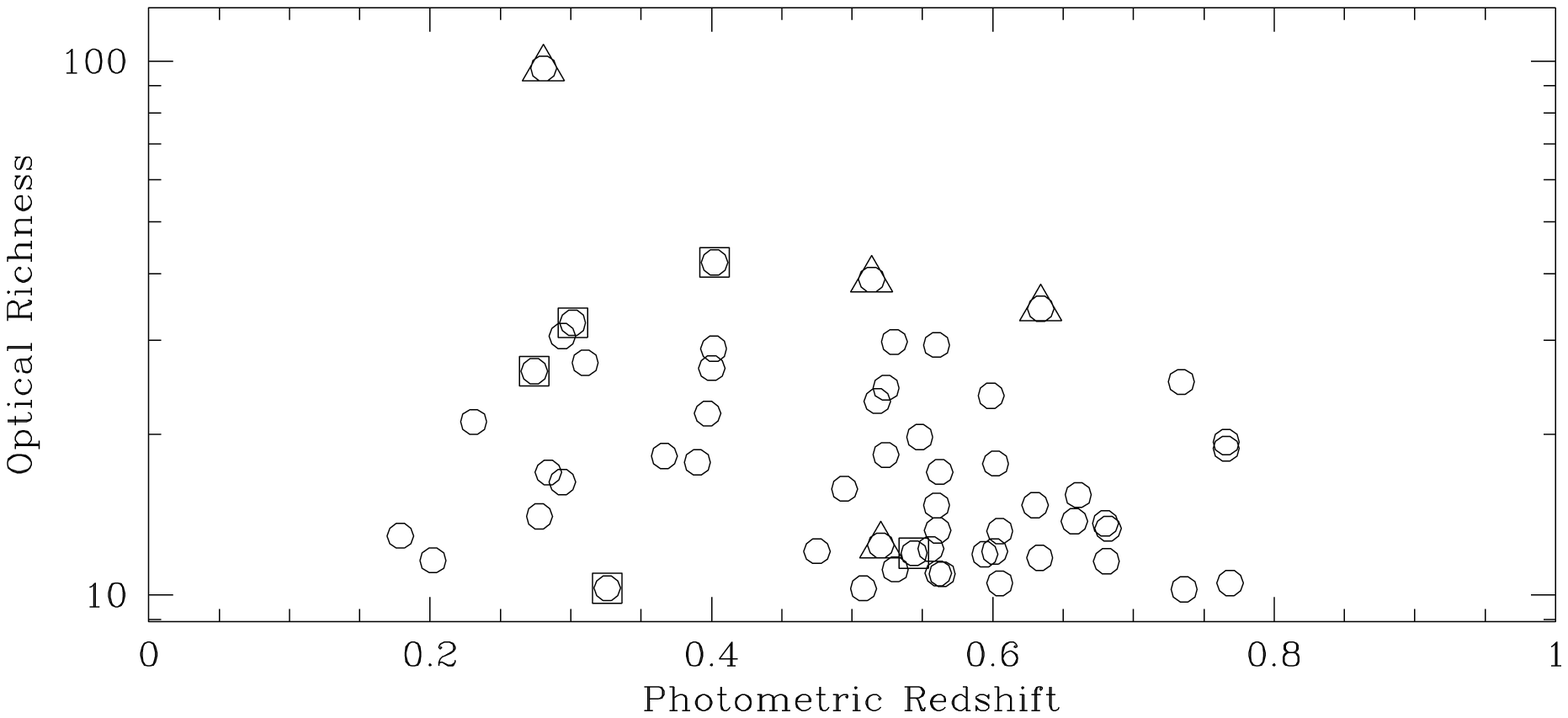}}}}
\figcaption{Photometric redshift versus richness plot of clusters detected by
  the CAMIRA algorithm (circles). Among them, clusters matched with the
  highest significance convergence peaks (weak lensing S/N $\ge$ 4.5) 
  are marked by triangles; lower significant ones (3.7 < S/N < 4.5) by squares.
\label{fig:z_richness}}
\vspace{0.3cm}

\subsection{Weak Lensing Mass Estimate}
The masses of the clusters are estimated from the tangential alignment of
the shear. The shear is azimuthally averaged over the successive annuli placed
with a logarithmic interval in radius of 0.2.
We fit the radial profile with a singular
isothermal sphere (SIS) model, (surface density $\propto \theta^{-1}$). 
In order to minimize dilution by the member galaxies of clusters (and
any intrinsic alignment signal)
we avoid the center;
the fitting was made from the radius of 1.8 arcmin to 12 arcmin. 

The tangential shear, $\gamma_{T}$, induced by an SIS model is 
\begin{equation}
\gamma_{T} = \frac{1}{\Sigma_{cr}}\frac{\sigma_{SIS}^{2}}{2G}\frac{1}{r}
\label{eqn:gt1}
\end{equation}
where $\Sigma_{cr}$, $\sigma_{SIS}$ and $r$ is the critical surface
density, the velocity dispersion of the SIS and radial distance from the
center;
\begin{equation}
\Sigma_{cr} = \frac{c^2}{4\pi G}\frac{D_S}{D_L D_{LS}}, \;\; r = D_L\theta .
\label{eqn:misc}
\end{equation}
Here, $D$ is angular diameter distance. 
By putting Eqn.~(\ref{eqn:misc}) into Eqn.~(\ref{eqn:gt1}), we have 
\begin{equation}
\sigma_{SIS}^2 =
\frac{c^2}{360}\frac{D_S}{D_{LS}}\gamma_{T}(\theta = 1 [deg])
\label{eqn:sigmasis}
\end{equation}
The mass of SIS within the radius, $r_{\Delta_c}$, is given by 
\begin{equation}
M(<r_{\Delta_c}) = \frac{2\sigma_{SIS}^{2}}{G}r_{\Delta_c}.
\label{eqn:m1}
\end{equation}
The mass is also described by adopting the critical over-density parameter,
$\Delta_c$,
with respect to the mean density, $\bar{\rho}(z)$, as
\begin{equation}
M(<r_{\Delta_c}) = \frac{4\pi}{3}r_{\Delta_c}^3 \bar{\rho}(z)\Delta_c(z) / \Omega_M   .
\label{eqn:m2}
\end{equation}
Eliminating $r_{\Delta_c}$ from Eqn.~(\ref{eqn:m1}) and Eqn.~(\ref{eqn:m2})
yields
\begin{equation}
\begin{split}
M(<r_{\Delta_c})  =  & 4.16 \times 10^{14}M_{\odot}h^{-1}(\sigma_{SIS}/1000[km/s])^3  \\
 & \times \left(\frac{1}{(1+z)^3}\frac{500}{\Delta_c}\right)^{\frac{1}{2}}, \\
\end{split}
\label{eqn:m3} 
\end{equation}
where $\bar{\rho}(z) = (1+z)^3\Omega_M\rho_{cr}^0$ is
adopted. We fit the data to obtain $\gamma_{T}(\theta=1 [deg])$ and 
the mass is estimated by using Eqn.~(\ref{eqn:sigmasis}) and
Eqn.~(\ref{eqn:m3}) where we employ the redshift distribution of the source
galaxies, $n(z_s)$, from \cite{lefevre13}
rather than adopting the photo-z estimated only from optical BVRz data to
  avoid possible systematic effects from photo-z outliers.
Note that the weak lensing mass inferred from SIS model does not
  differ from   what is obtained from NFW model in our fitting
  range. The difference depends on the mass but it is  less than
10 \% in most cases, which is rather smaller than the statistical
error.

Based on the comparison of the CFHT MegaCam data with the GREAT simulation,
\cite{miller13} argued that {\it lens}fit slightly underestimates the
multiplicative factor, $m$, of the shear especially when the signal to
noise ratio of the objects, $\nu$,  is low; $\nu < 20$ whereas the
underestimate becomes less than 5 \% when $\nu > 30$. Although we have
not completed a comprehensive comparison of HSC data with the
simulation the behavior of {\it lens}fit on HSC data should not be
significantly different from that of CFHT MegaCam because the image
quality and pixel scale are similar. Therefore, we should have a level
of roughly 5 \% underestimate of shear at maximum which results in 
a roughly 7 \%
underestimate in the mass evaluation. Since this is small
compared with the statistical error, we will not deal with the
systematic error explicitly in this paper.

Table~\ref{tbl:masscomparison} shows the result of the comparison of 
our mass estimate of A781 components with the values in the literature
\citep{wittman14}. Despite the totally independent observation and
the analysis, the agreement is encouraging and implies some progress
in the convergence of weak lensing data analysis techniques.

\begin{table}
\caption{Comparison of the mass estimate
  ($M_{200}/10^{14}M_{\odot}$). The error is 1$\sigma$. Note that West
cluster mass is estimated at X-ray peak whereas all the other mass are
estimated at the weak lensing peak position.}
\label{tbl:masscomparison}
\begin{center}
\begin{tabular}{lcc}
\hline
A781 & This Work & \cite{wittman14} \\ \hline
Main   & 7.0$^{+1.8}_{-1.6}$ & 6.7$^{+1.4}_{-1.3}$ \\
Middle & 4.6$^{+1.4}_{-1.2}$ & 4.3$^{+1.6}_{-1.2}$ \\
East   & 4.8$^{+1.6}_{-1.3}$ & 2.8$^{+1.9}_{-1.2}$ \\ \hline \hline
West   & 1.2$^{+0.7}_{-0.5}$ & 2.7$^{+1.5}_{-1.0}$ \\ 
\hline
\end{tabular}
\end{center}
\end{table}

\vspace{0.3cm}
\centerline{{\vbox{\epsfxsize=8.5cm\epsfbox{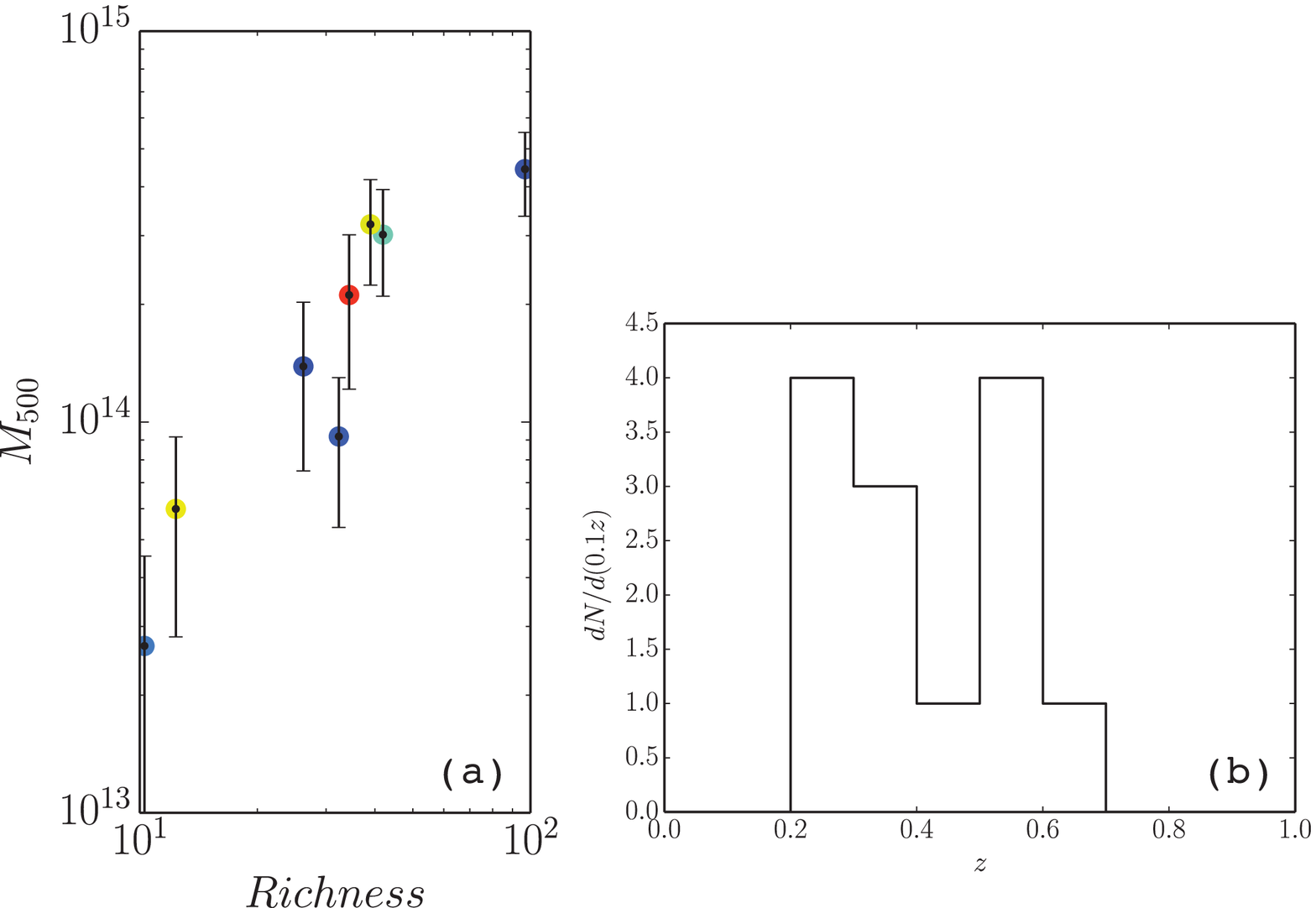}}}}
\figcaption{Richness versus $M_{500_{c}}$ of the shear selected samples that
  have optical counterparts detected by the CAMIRA algorithm (a). Redshift
  distribution of all identified shear selected clusters (b).
\label{fig:richnessmass.nz}}
\vspace{0.3cm}

Fig.~\ref{fig:richnessmass.nz} (a) shows the relation between the richness
and cluster virial mass, $M_{500}$ of the optically selected cluster samples 
that
have counterparts with the convergence peaks. The correlation is clearly
seen and the slope is roughly 1$\sim$ 1.5, which agrees nicely with the
slope found in \cite{oguri14} 
where the richness-mass relation was derived
from stacked weak lensing analysis with the CFHTLenS shear catalogue.
This further supports
the reality of the cross-match of our lensing peaks and the 
optically selected clusters. 

In Fig.~\ref{fig:richnessmass.nz} (b), we show the redshift distribution
of the identified shear selected clusters on
Table\ref{tab:shear_selected_cluster}. 
We see two spikes around z of 0.3 and 0.5 which clearly indicates 
that this narrow 2.3 deg$^2$ field is populated by large scale
structure at these redshifts; this has been mentioned already
by \cite{kubo09}. We need a significantly wider field of view to 
overcome the local variance and to make cosmological arguments from
the redshift distribution of clusters.

\subsection{Peak Count and Comparison with the Theoretical Estimates}
The observed area which overlaps with DLS amounts to 2.3 deg$^2$, in which we
found eight significant peaks whose S/N exceeds 4.5. Even if we drop
Peak ID 6 from the list, which may be substructure within the A781 main
cluster, seven peaks still remain. \cite{hu03} estimated the cosmological
variance in cluster samples and suggested that the
variance exceeds the shot noise when the mass of clusters becomes less
than $\sim 3\times 10^{14} M_{\odot}$ with a weak dependence of
the threshold mass on the survey volume. This is exactly the mass 
range that we are working on. Therefore, statistical arguments require
comparison with cosmological simulations as we will see below.

\vspace{0.3cm}
\centerline{{\vbox{\epsfxsize=8.5cm\epsfbox{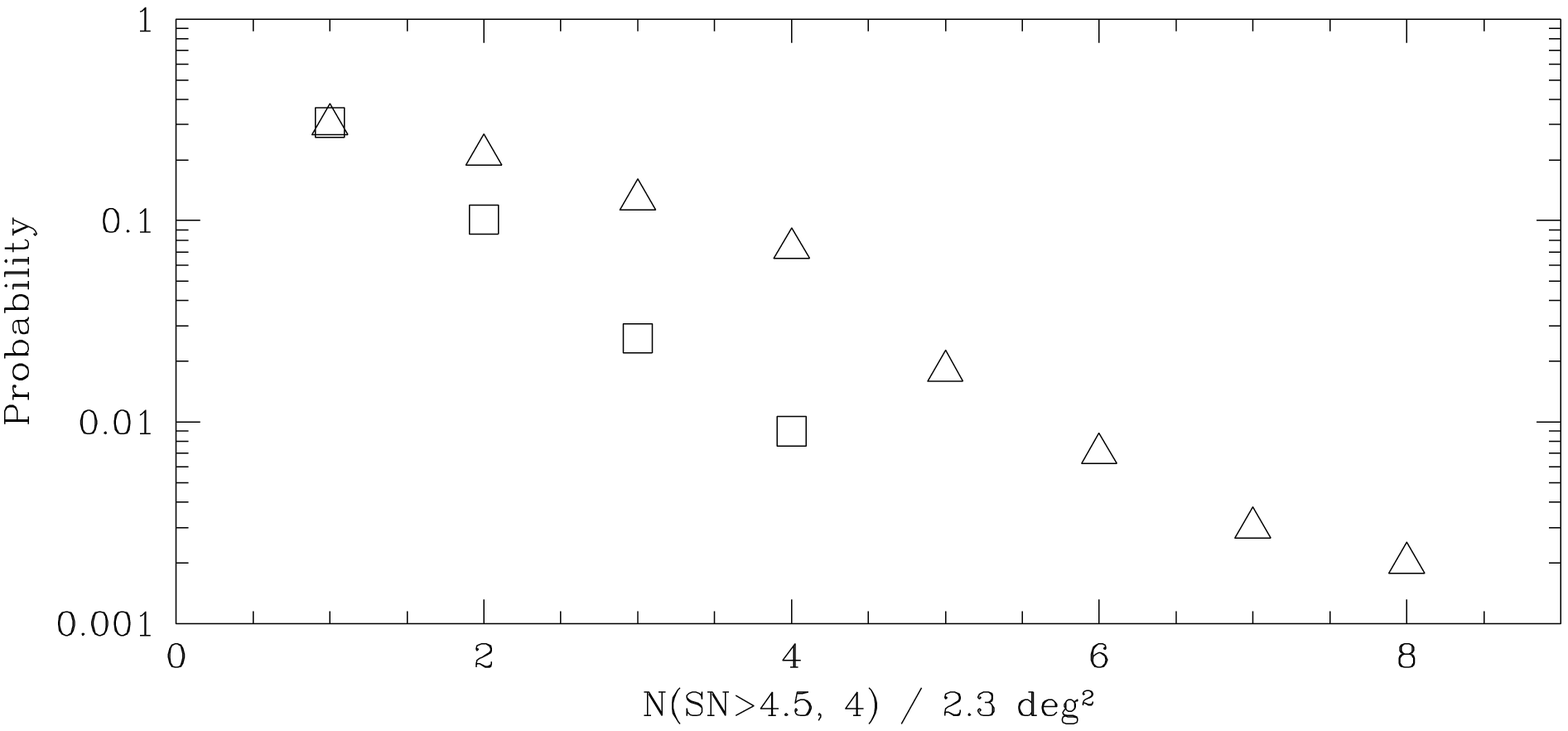}}}}
\figcaption{Probability distribution of the number of the peaks under a
  given S/N threshold on a 2.3 deg$^2$ wide field. Squares for S/N > 4.5
  and triangles for S/N > 4.0. 
\label{fig:hamana12}}
\vspace{0.3cm}

\cite{hamana12} calculated the number of peaks on the weak
lensing convergence map using a large set of gravitational lensing
ray-tracing simulations which are detailed in \cite{sato08}.
Following the work, we made 1000 realizations to evaluate the sample
variance. In making the mock weak lensing convergence map, we added a random
galaxy shape noise to the lensing shear data. The root-mean-square
(RMS) value of the random galaxy shape noise was set so that 
the observed galaxy number density and RMS of galaxy ellipticities
are recovered. We adopted a fixed source redshift of $z_s=1.0$.
\cite{lefevre13} estimated the redshift distribution of magnitude 
limited samples taken from VIMOS VLT Deep Survey, and reported the mean
redshift of $\langle z\rangle=0.92$ for a sample of $17.5 \leq I_{AB}
< 24$ and $\langle z\rangle=1.15$ for a sample of $17.5 \leq I_{AB} <
24.75$. Therefore, it would be appropriate to set $z_s=1$ for
our galaxy sample of $i_{AB} < 24.5$.

The expected peak count of S/N > 4.5 is  0.61 on 2.3 deg$^2$. 
In Fig.~\ref{fig:hamana12}, we show the probability distribution 
of the number of peaks under a  given S/N threshold on a 2.3
deg$^2$ wide field.  As is shown in square symbols, the maximum number of peaks
reached  in the realization is 4 (9/1000) when the S/N = 4.5. This
means that we have practically no chance to have seven or eight
significant peaks on 2.3 deg$^2$ field. Is this a challenge against
the current CDM based cosmology?

We note that the sensitivity of the  number of  peak to the S/N value
is quite high, reflecting the steepness of mass function at the high
mass end. So we experimentally lower the S/N down to 4 and examine the
statistics. The mean number of peaks is 1.6 and the maximum number of
peaks is eight in two realizations out of 1000 (0.2\%, see triangles
in Fig.~\ref{fig:hamana12}). This still does not reconcile the gap
between the observation and the prediction.

In the mean time, \cite{hamana12} had adopted cosmological simulation
which used 3rd year WMAP result \citep{wmap3}. WMAP3 is known to have
yielded a relatively low $\sigma_8$ of  0.76. If we adopt the recent
Planck result \citep{planck15}, $\sigma_8 = 0.83$, 
the expected cluster count  becomes a factor of 5.26 higher than
WMAP3. This relaxes the tension dramatically and now 
what we observed is not extremely unlikely; the chance to obtain
more than 8 peaks of S/N > 4.5 is  3.7 \%. 
One thing  that we could suggest here is that our peak count strongly
favours the recent Planck result. 

\subsection{Stellar Mass Fraction in Clusters of Galaxies}

\vspace{0.3cm}
\centerline{{\vbox{\epsfxsize=8.5cm\epsfbox{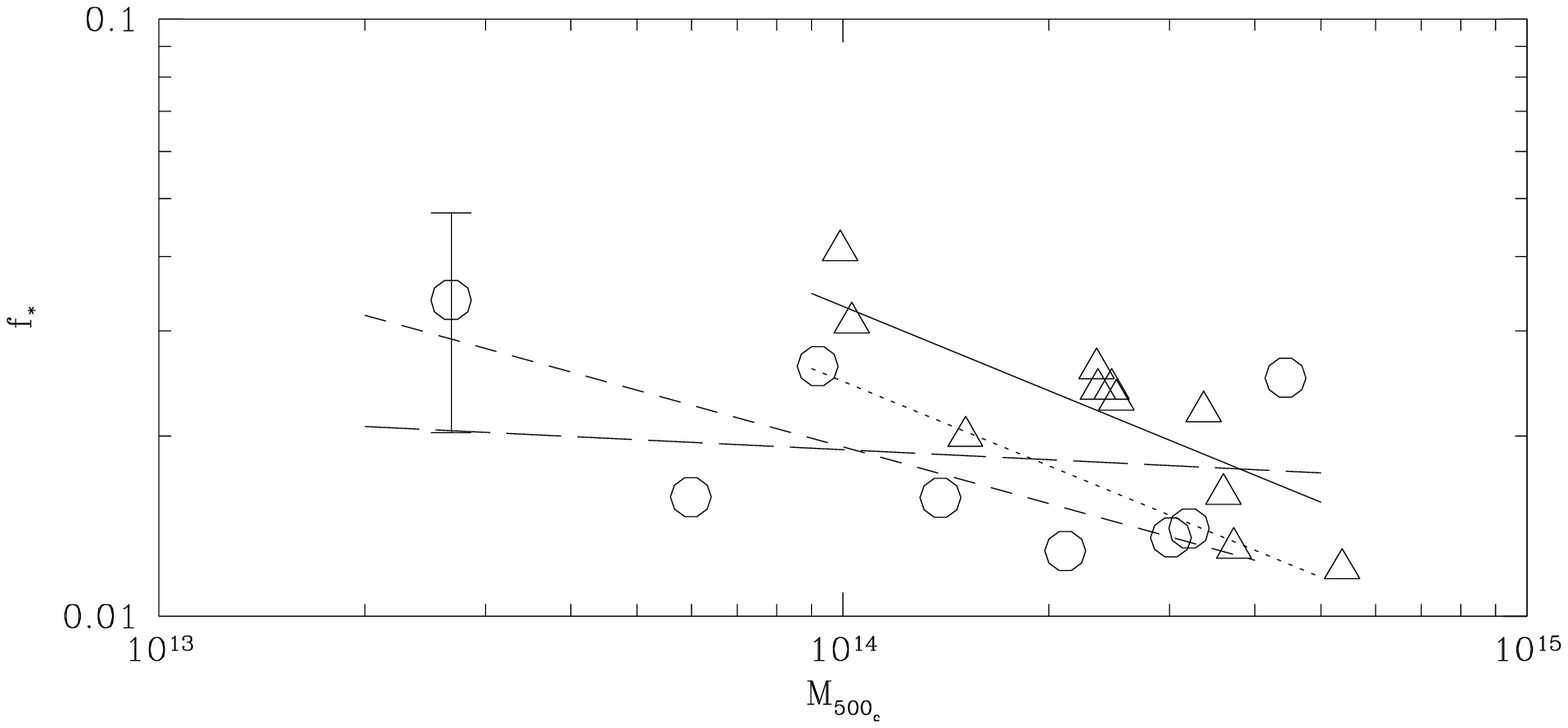}}}}
\figcaption{Fraction of stellar mass over the  halo mass of the shear selected
  clusters (Circles). 
The dark matter virial halo mass, $M_{500}$, is estimated by putting
$\Delta_c = 500$ in Eqn.~(\ref{eqn:m3}).  
Triangles are from \cite{gonzalez13} Table 6 and 7, and include the
stellar mass associated with intracluster light (ICL). 
Solid line is the best fit to the triangle data. When they exclude
the contribution of ICL outside of 50 kpc, the result is given by
the dotted line. Typical error of 40 \% is presented on the
leftmost data point. Short-dash-line is the best fit power function for
all the clusters and long-dash-line for the clusters excluding the
most massive A781 Main. 
\label{fig:fsMdm}}
\vspace{0.3cm}

We now examine the ratio of stellar mass to halo mass, $f_s =
M_s/M_{halo}$, of our samples. In the mass range that we probe, $f_s$ 
is reported to decrease as the halo mass increases, suggesting that
star formation is less efficient in larger halos. This can be
mostly explained by the inefficiency of the cooling process inside larger
halos. \cite{gonzalez13} presented one of the most recent results based on
new halo mass estimates from XMM-Newton X-ray data. They claimed that
the stellar baryon mass well compensates the shortage in the baryon
budget and that the sum of the stellar baryon and the baryons in the form
of gas almost reaches
to the universal value estimated by WMAP and Planck. They also
made a comparison of different observational results \citep{lin03,
gonzalez07, andreon10, lin12, leauthaud12}. They found that the stellar
mass fractions reported in other works are generally lower than
\cite{gonzalez13}. However, they claimed that if the mass in the 
intra-cluster light
(ICL) is considered, the discrepancy among the observations is
minimized.

Fig.~\ref{fig:fsMdm} shows the $f_s$ versus the halo
mass of our samples (circles). The mass is estimated by weak
lensing as explained previously and the mass range of the samples is
wider than that of previous small samples.
The stellar mass $M_s$ shown in Table 2 is the total stellar mass
integrated by convolving spatial filter as shown in Fig. 2 of Oguri
(2014). In order to estimate the stellar mass fraction accurately, we
convert this stellar mass to the stellar mass within $r_{500_c}$ as
follows. We assume that the stellar mass density profile follows an
NFW profile with the concentration parameter as a function of
halo mass of Duffy et al. (2008).
For each halo with the mass $M_{500_c}$, we derive a conversion factor
from $M_s$ in Table 2 to the stellar mass within $r_{500_c}$ by
convolving the projected NFW profile with the spatial filter of Oguri
(2014) and estimating the ratio of the total mass with the spatial
filter to $M_{500_c}$. We find that the conversion factor is $\sim 1$
for massive halos with $M_{500_c}\sim 5\times 10^{14}M_\odot$, and
$\sim 0.5$ for less massive halos with $M_{500_c}\sim 5\times
10^{13}M_\odot$. We note that this correction also takes account of
the 3D de-projection, i.e., properly removes member galaxies outside
$r_{500_c}$ projected along the line-of-sight.
Thus our result should be compared with the so-called `3D' stellar mass
result in \cite{gonzalez13} that incorporates the de-projection.
The error in $M_{halo}$ estimate is roughly 30
\%. We also expect a large scatter of $\sim 30$\% in $M_s$ for a given
halo mass largely originating from the Poisson noise in the numbers of
cluster member and background galaxies.
Therefore, we expect as large as 40 \% level error in estimate of the 
individual $f_s$. The typical error bar is presented on the leftmost 
data point in Fig.~\ref{fig:fsMdm}.

Compared with \cite{gonzalez13} results (triangles and the solid line in
Fig.~\ref{fig:fsMdm}),  $f_s$ of this work is lower on average over
our mass range. We did not attempt to include the contribution from 
the ICL but according to \cite{gonzalez13}, the ICL contribution 
is $\sim$ 25 \% (dotted line in the figure) which
cannot totally explain the discrepancy.  Our result agrees better with
lower $f_s$ results estimated by  Halo Occupation Distribution 
(HOD) method in \cite{leauthaud12} and favors the argument 
that missing baryon problem has not yet been resolved in
this mass range. 
We further note that \cite{gonzalez13} assumed a stellar
mass to luminosity ratio that is on average slightly lower
than the Salpeter initial mass function adopted in this paper.
Our stellar mass estimates are therefore larger than their estimates
by $\sim10\%$ which makes the discrepancy of the results larger by this
factor. 

The decrease rate of $f_s$ with increasing halo mass is an 
interesting observable because it reflects how the clusters are
formed. Theoretical predictions and the recent N-body simulations
coupled with semi-analytic models disfavor the steep slope in the
framework of $\Lambda$CDM cosmological model. This is because
the hierarchical clustering predicts that low mass halos with a
certain $f_s$ would be assembled together and become a larger halo
with a comparable $f_s$. \cite{balogh10} suggested that the log-log slope of
$\sim$ --0.3 would be the upper limit to be consistent with their
simulations in the $\Lambda$CDM cosmology.
The slopes that have been observed and were compiled in
\cite{gonzalez13} span over --0.3$\sim$--0.6 and the slope of --0.45 is
presented by the author's data which is slightly steeper than the
theoretical preference. In our case, the slope is slightly shallower
than the data of \cite{gonzalez13}; --0.32 when we exclude A781 Main
(long-dash-line) and very shallow --0.05 (short-dash-line) when we
consider all the clusters. It is, however, difficult to make really
qualitative arguments here due to the large error for the limited
number of clusters in this work. More data are certainly cried
  for and the  on-going HSC legacy survey will provide ideal sources for the future
  studies. 

\section{Conclusion}
We show the results of a weak-lensing cluster search on  
2.3 square degrees of HSC commissioning data in the Abell 781 field with 
1.6 hours of exposure. The data are excellent, with very good image
quality; some slight astigmatism was found which was traced to
a small miscollimation (since corrected) but was not difficult
to correct because of the low order spatial variation.

Clusters were searched for on a high resolution convergence map generated
froma these data.
We see very good agreement with the previous results 
in mass measurements in this field made by Wittman et al (2014) except 
for one peak,
corresponding to Abell 781 West. Clusters of galaxies were searched 
for in optical data using
CAMIRA algorithm \citep{oguri14}. This cluster list was compared with
the locations of the peaks in the convergence map. 
There is only one significant peak which we cannot judge the
reality because it is outside of DLS field where no multi-color data
is available. All the other peaks of S/N > 4.5 are physically real.
This demonstrates the reliability of the convergence map generated
from these even early HSC data. The number of observed 
clusters at this level is significantly larger than the predicted 
average number in this field size expected from theoretical calculations
based on WMAP-3 results. 
This result, however, is extremely sensitive to the value of $\sigma_8$
in the theoretical predictions, and is not unlikely if we adopt the recent
Planck cosmology with its somewhat higher value of $\sigma_8 = 0.83$. 


The stellar mass fraction in our sample
is systematically lower than one
of the other most recent results and is more consistent with the earlier values
estimated by use of the HOD statistical formalism. Our
result thus favors the argument that baryons are still missing in this mass
range. The decrease of the stellar mass fraction with increasing halo
mass is slightly shallower than the previous work, and is more
consistent with current (though still very uncertain) simulations.

Because of both the limited sample size and the statistical errors in
our masses, the results for the stellar mass fraction are not strictly
conclusive, but {\it are} strongly suggestive.

So even though the results for number density and stellar mass
fraction from this very small sample are not conclusive, samples
not enormously larger than this with this instrumentation will be sufficient
to provide excellent results on these important quantities, and a large
survey is underway which will provide these data.

In this work, we were at least able to demonstrate that cluster
identification, redshift estimates, and mass estimates can 
be obtained by multi-band optical imaging data with the newly 
developed Hyper Suprime-Cam camera through weak lensing and cluster
finding techniques. HSC has uniquely combined features of 
wide field, large aperture and superb image quality and 
the data from the currently on-going HSC legacy survey is very promising
for these and many other cosmological investigations.

\acknowledgments
\subsubsection*{Acknowledgments}
We are very grateful to all of Subaru Telescope staff. 
This work was supported in part by Grant-in-Aid for
Scientific Research from MEXT (18072003), the JSPS (26800093) and
World Premier International Research Center Initiative (WPI
Initiative), MEXT. This paper makes use of software developed for the
Large Synoptic Survey Telescope. We thank the LSST Project for making
their code available as free software at http://dm.lsstcorp.org.

\end{document}